# Spin-Axis-Layer Locking for Intrinsic Bipolar Altermagnetic Semiconductors: Proof-of-Concept in Bilayer $CuBr_2$

Wei Ma, Dengpan Ma, Zhiheng Lv, Zhifeng Liu[*]

*Research Center for Quantum Physics and Technologies & Inner Mongolia Key Laboratory of Microscale Physics and Atom Innovation, School of Physical Science and Technology, Inner Mongolia University, Hohhot 010021, China*

**Abstract**

Electrical control of spin and magnetic sublattice degrees of freedom is essential for multifunctional and low-power spintronic devices. Bipolar altermagnetic semiconductors (BAMSs)—characterized by opposite spin polarizations at the valence and conduction band edges—offer such control, yet known systems require external strain and sizable valley polarization for gate-tunable switching. Here, we propose a universal spin-axis-layer locking (SALL) paradigm to overcome these limitations. By stacking two quasi-1D ferromagnetic monolayers with a 90° twist, the bilayer reconstructs altermagnetic symmetry, yielding an intrinsic BAMS state where carrier spin is locked to specific layers and transport directions. Using synthesized $CuBr_2$ monolayers as proof-of-concept, we demonstrate via first-principles calculations a robust BAMS state. Electrostatic gating enables simultaneous, reversible switching of carrier type, spin, and active layer, generating fully spin-polarized axial charge currents and directionally controllable pure spin currents with near-unity charge-to-spin conversion efficiency. This SALL model establishes a versatile, strain-independent strategy for advanced all-electrical altermagnetic devices.

Electrical control of reversible, highly spin-polarized transport is a cornerstone of spintronics [1-3]. Bipolar magnetic semiconductors (BMSs) [4-6], featuring opposite spin polarizations at the valence and conduction band edges, enable gate-tunable switching between fully spin-polarized electron and hole currents. This capability has stimulated extensive theoretical [6-11] and experimental efforts, exemplified by systems such as VNbRuAl [12] and the MnPX$_3$ (X = S, Se) family [13, 14]. However, ferromagnetic (FM) BMSs possess finite net magnetic moments, which generate stray fields and increase susceptibility to perturbations, thereby limiting device integration and stability [15]. These drawbacks have driven the search for bipolar antiferromagnetic (AFM) semiconductors (BAFSs) [16], explored in platforms ranging from asymmetric transition-metal compounds [17] and surface-functionalized MXenes [18] to gated van der Waals (vdW) A-type antiferromagnets [19-22]. Despite their zero net magnetization and suppressed stray fields, conventional collinear antiferromagnets preserve strict spin degeneracy enforced by crystal symmetry, precluding intrinsic spin splitting [23, 24]. As a result, achieving sizable spin polarization typically requires strong external perturbations, hindering the realization of FM-like spin splitting.

Recently discovered altermagnets [24-27] emerge as a distinct collinear magnetic phase that combines the large, FM-like spin splitting with the zero net magnetization of antiferromagnets, offering a route to overcome these long-standing bottlenecks. Leveraging this concept, we recently proposed the paradigm of bipolar altermagnetic semiconductors (BAMSs), identifying monolayer Ti$_2$Br$_2$O as a promising candidate [28]. In that system, however, uniaxial strain is required to break the diagonal mirror symmetry and induce valley splitting, which in turn locks the spin orientation at the band edges to enable bipolar-tunable transport. Such reliance on external symmetry breaking complicates device integration and precludes purely electrical control. A fundamental question therefore remains: can an intrinsic BAMS be realized to achieve fully spin-polarized transport with reversible polarity, independent of external strain or valley polarization?

We address this challenge by leveraging the spin-axis dynamic locking (SADL) effect [29] in 2D altermagnets, which originates from a unique "tent-state" electronic structure characterized by momentum-dependent spin splitting, alternating spin-polarized flat bands, and orthogonal Fermi-surface contours [29]. Consequently, carrier transport becomes highly anisotropic and spin selective: conduction along one principal crystal axis selects a specific spin channel, while the orthogonal axis selects the opposite. This deterministic locking between spin orientation and macroscopic current direction provides a robust route to realizing BAMS functionality without the need for symmetry-breaking strain or valley polarization.

Leveraging this mechanism, we incorporate the layer degree of freedom to formulate a spin-axis-layer locking (SALL) mechanism as a structural strategy for intrinsic BAMSs. Inspired by orthogonally twisted anisotropic bilayers [30, 31] and cross-chain magnetic superlattices [32], we construct the SALL model by vertically stacking two identical quasi-1D FM monolayers with a 90° relative rotation. This cross stacking stabilizes A-type AFM alignment while reconstructing the required altermagnetic symmetry at the bilayer level. Using the synthesized chain-like $CuBr_2$ monolayer [33] as a proof-of-concept, we demonstrate that the bilayer embodies this SALL paradigm, where carrier spin is intrinsically locked to both its active layer and transport direction. First-principles calculations reveal that electrostatic gating enables the simultaneous, reversible switching of carrier polarity, spin, and transport axes. This functionality allows for the generation of uniaxial fully spin-polarized currents and pure spin currents with near-unity efficiency. Furthermore, the spatial segregation of spin channels into distinct layers provides a unique pathway for independent spin extraction via lateral contact geometries. By validating the SALL mechanism, this work establishes a generalizable framework for the gate-tunable control of altermagnetic semiconductors, providing a superior, strain-free platform for highly integrated, low-power spintronic devices.

**A. Theoretical Model of Spin-Axis-Layer Locking (SALL).** To realize an intrinsic BAMS state without external stimuli, we propose the SALL paradigm. As illustrated in Fig. 1a, the model comprises two identical, quasi-1D FM monolayers, in which electron hopping is confined to a single axis with negligible transverse dispersion. Stacking the two layers with a 90° relative rotation—forming bottom (BL) and upper

(UL) layers—yields a cross-stacked A-type AFM bilayer. The 90° rotation is crucial for reconstructing altermagnetic symmetry. While an isolated quasi-1D FM monolayer possesses only $C_2$ rotational symmetry, the cross-stacked bilayer establishes an effective $C_{4z}$ equivalence between layers. Although the interlayer displacement and orthogonal chain orientations preclude a strict $C_{4z}$ space-group symmetry, the system remains invariant under the combined spin-group operation [$C_2 \parallel C_{4z}$] [24, 34], where $C_{4z}$ transforms the lattice geometry and hopping anisotropy, and $C_2$ acts in spin space to reverse the AFM spin polarization.

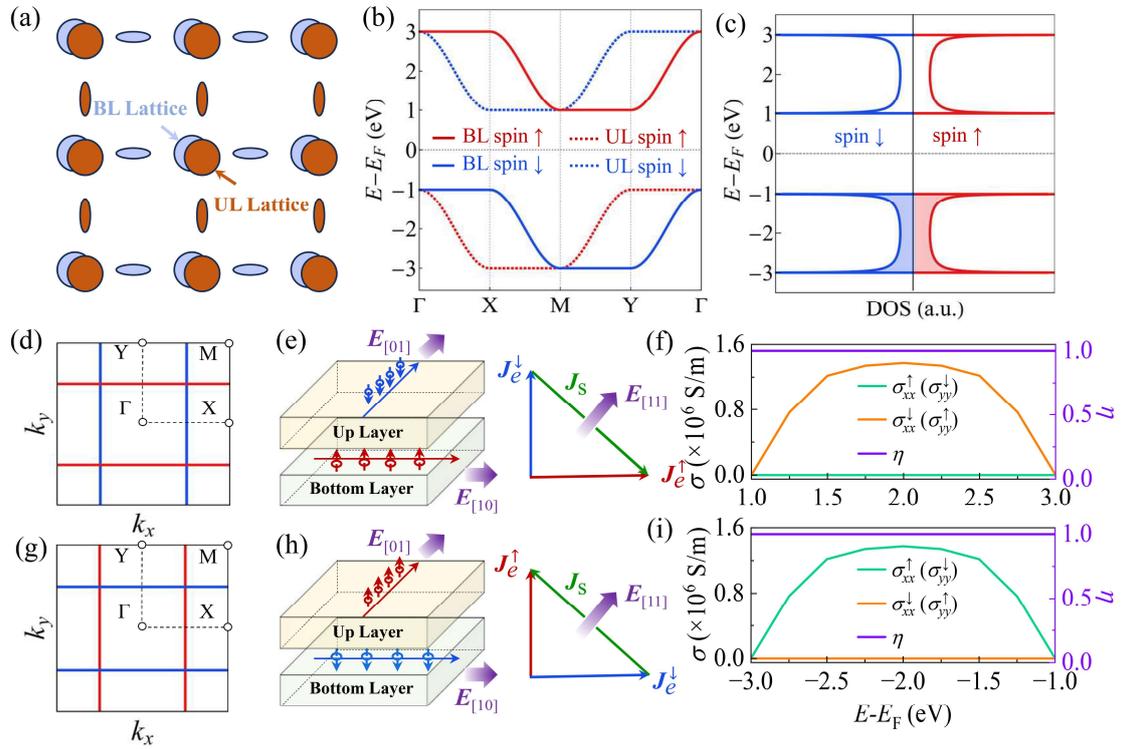

**Fig. 1.** Schematic illustration of the SALL model for intrinsic BAMSs. (a) Schematic of the lattice model comprising two perpendicularly stacked quasi-1D FM monolayers (UL: upper layer; BL: bottom layer). (b) Spin- and layer-resolved band structure and (c) the corresponding DOS. Orthogonal Fermi-surface contours for the conduction (d) and valence (g) bands. Schematics of the SALL effect for (e) electron- and (h) hole-doped regimes. Energy-dependent conductivities and spin polarization ($\eta$) for electrons (f) and holes (i).

The low-energy physics is captured by a minimal tight-binding (TB) Hamiltonian. Owing to the large vdW gap and the orthogonality of the 1D conducting channels, interlayer hybridization is negligible, and the Hamiltonian decomposes into two decoupled subsystems, $H = H_{BL} \oplus H_{UL}$. For each layer $l$ = BL, UL, the Hamiltonian includes an on-site energy $\varepsilon$, the nearest-neighbor hopping $t$ along the chain, and an exchange splitting $\Delta$. The real-space Hamiltonian reads:

$$H_l = \varepsilon \sum_{i,\sigma} c^\dagger_{i,l,\sigma} c_{i,l,\sigma} - t \sum_{\langle i,j \rangle,\sigma} c^\dagger_{i,l,\sigma} c_{j,l,\sigma} - s_l \Delta \sum_{i,\sigma} \sigma c^\dagger_{i,l,\sigma} c_{i,l,\sigma} .$$

Here, $c^\dagger_{i,l,\sigma}$ and $c_{i,l,\sigma}$ are the creation and annihilation operators, and $\langle i,j \rangle$ denotes nearest neighbors along the $x$-axis (BL) or $y$-axis (UL). The spin index $\sigma = \pm 1$ corresponds to ↑/↓, and $s_l = \pm 1$ (with $s_{BL}=+1$, $s_{UL}=-1$) enforces the A-type AFM alignment, leading to opposite exchange splitting in the two layers. In momentum space, under the basis $\Psi_k = (c_{k,BL,\uparrow}, c_{k,BL,\downarrow}, c_{k,UL,\uparrow}, c_{k,UL,\downarrow})^T$, the Hamiltonian becomes diagonal: $H(k) = \text{diag}(\varepsilon_{BL,\uparrow}, \varepsilon_{BL,\downarrow}, \varepsilon_{UL,\uparrow}, \varepsilon_{UL,\downarrow})$, with $\varepsilon_{BL,\uparrow/\downarrow} = \varepsilon \mp \Delta - 2t \cos k_x$ and $\varepsilon_{UL,\uparrow/\downarrow} = \varepsilon \pm \Delta - 2t \cos k_y$ (see Supporting Information (SI) section S1). The resulting band structure (Fig. 1b) is fully spin- and layer-resolved. For the BL ($x$-direction), spin-up states form the valence band and spin-down states the conduction band, whereas this correspondence is reversed in the UL ($y$-direction). Coupled with the $[C_2 \| C_{4z}]$ symmetry, these band characteristics yield a momentum-dependent spin splitting, $\Delta E(k) \propto \cos k_x - \cos k_y$ (see Sec. S1, SI), exhibiting alternating positive and negative signs across the Brillouin zone ($d$-wave state). The density of states (Fig. 1c) exhibits pronounced spin splitting and van Hove singularities, characteristic of quasi-1D systems.

The Fermi surfaces consist of orthogonal open 1D contours for both conduction (Fig. 1d) and valence (Fig. 1g) bands. This leads to highly anisotropic, spin-selective transport. In the electron-doped regime (Fig. 1e), spin-up electrons conduct along $x$ in the BL, while spin-down electrons conduct along y in the UL. An in-plane electric field therefore generates a fully spin-polarized current confined to a single layer and axis. In the hole-doped regime (Fig. 1h), the spin-layer correspondence is exactly reversed. The calculated conductivities confirm strictly uniaxial, fully spin-polarized transport ($\eta$ = 100%) for both electrons (Fig. 1f) and holes (Fig. 1i). This threefold locking among spin, layer, and transport direction defines the SALL effect.

A distinct consequence arises under a diagonal in-plane electric field (θ = 45°).

Owing to the orthogonal, layer-resolved transport channels established above, the longitudinal charge currents combine along the field direction, whereas along the transverse (−45°) direction the charge components cancel and the opposite spin polarizations add, yielding a pure spin current with 100% charge-to-spin conversion efficiency. Upon electrostatic switching between electron- and hole-doped regimes, the spin-layer correspondence is reversed, and the transverse pure spin current correspondingly flips its polarization under the same field. This enables fully electrical, on-demand control of carrier type, active transport channel, and the direction of pure spin currents.

**B. The Quasi-1D Building Block: Monolayer CuBr$_2$.** To realize the proposed SALL model, a candidate material should simultaneously exhibit intrinsic ferromagnetism, bipolar spin polarization, and strongly anisotropic 1D transport. Based on first-principles calculations (see SI section S2 for method details [35-45]), we identify monolayer CuBr$_2$ as an ideal proof-of-concept platform. Recent experiments confirm that layered vdW CuBr$_2$ can be synthesized and exfoliated down to the atomically thin limit [33]. The monolayer CuBr$_2$ crystallizes in the C2/*m* space group (Fig. 2a), featuring parallel quasi-1D chains propagating along the *y* axis. Within each chain, a linear Cu backbone is bridged by Br pairs, forming edge-sharing CuBr$_4$ ribbons that promote strong intra-chain electronic hybridization. Conversely, neighboring chains interact mainly via weak vdW forces. Our optimized lattice (a = 7.111 Å, b = 3.515 Å) yield an intra-chain Cu-Cu distance of 3.515 Å, which is significantly shorter than the inter-chain separation (3.97 Å). These parameters agree well with X-ray diffraction data of exfoliated CuBr$_2$ flakes [33]. The large transverse spatial gap and the absence of covalent bridging quench inter-chain electronic coupling. This quasi-1D structural motif is vividly corroborated by the simulated scanning tunneling microscopy (STM) image (Fig. 2b), providing direct physical justification for the decoupled orthogonal hopping assumed in our SALL Hamiltonian.

First-principles calculations confirm that the magnetic ground state of monolayer CuBr$_2$ is intrinsically FM (see SI section S2). Due to the quasi-1D motif, inter-chain magnetic exchange is relatively weak, yielding a Curie temperature ($T_C$) of approximately 38.6 K via Monte Carlo simulations (see Fig. S2 in SI section S3), being comparable with that (45 K) of the synthesized FM CrI$_3$. While cryogenic, this $T_C$ is comparable to that of prototypical 2D magnets like CrI$_3$ (45 K), making it entirely sufficient to serve as a proof-of-concept platform for the SALL mechanism.

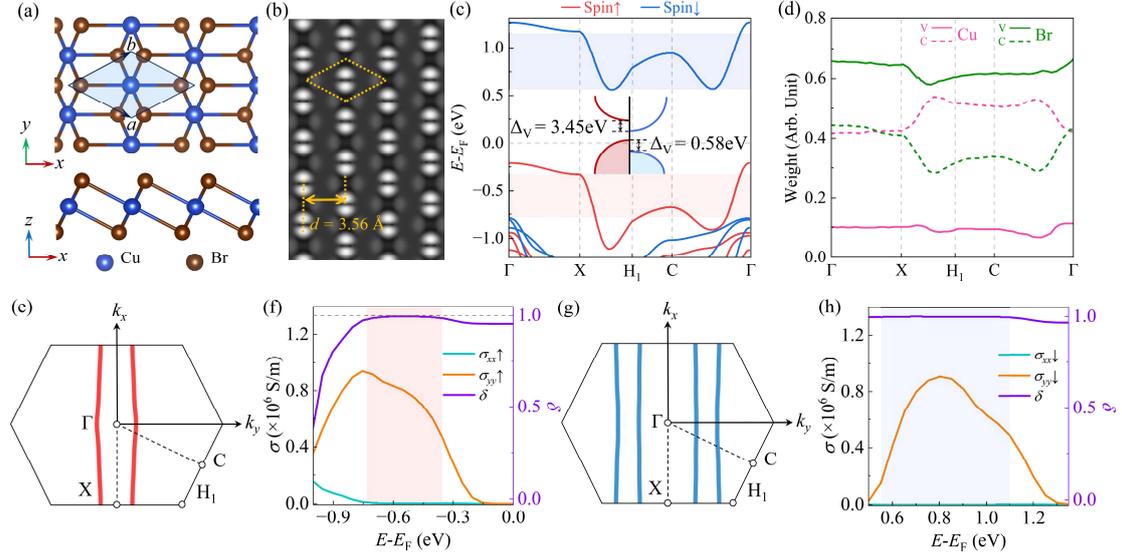

**Fig. 2.** Characterization of the quasi-1D FM CuBr$_2$ monolayer building block. (a) Crystal structure and (b) simulated STM image. (c) Spin-resolved band structure, with the inset illustrating the ideal BMS nature. The shaded regions highlight the broad energy windows that host open 1D Fermi surfaces. (d) Atom-resolved wave-function weights for the topmost valence (V, solid lines) and bottommost conduction (C, dashed lines) bands. 2D Fermi surface contours for the (e) valence and (g) conduction bands. Calculated energy-dependent conductivities ($\sigma_{xx}$, $\sigma_{yy}$) and the directional polarization ratio ($\delta$) for the (f) hole- and (h) electron-doped regimes. The corresponding shaded areas in (f) and (h) demonstrate the robust energy ranges where extreme spatial anisotropy ($\delta \approx 1$) is maintained.

Electronically, monolayer CuBr$_2$ behaves as an ideal FM BMS. Its spin-resolved band structure (Fig. 2c) exhibits a sizable global bandgap with reversed spin polarizations at the frontier band edges: the topmost valence band (TVB) is exclusively spin-up, while the bottommost conduction band (BCB) is fully spin-down. The giant exchange splittings ($\Delta_V$ = 3.45 eV and $\Delta_C$ = 0.58 eV) robustly prohibit thermally induced spin-flip scattering. Atomic projections (Fig. 2d) reveal that the TVB is predominantly derived from Br atoms, whereas the BCB exhibits stronger Cu hybridization. This spatial decoupling of spin-polarized electrons and holes inherently suppresses recombination, yielding extended carrier lifetimes that are highly advantageous for spintronic device integration [3, 46]. More fundamentally, orbital-projected band analysis (see Fig. S3 in SI section S4) reveals that these low-energy states are overwhelmingly dominated by the Cu-$d_{yz}$ and Br-$p_y$/$p_z$ orbitals, with an absence of $x$-oriented components (e.g., Br-$p_x$). Since these active orbitals align within the intra-chain bonding plane ($yz$-plane), their wavefunctions have practically zero extension along the inter-chain $x$-direction. This severe orbital confinement provides the microscopic origin for the quenched transverse hopping.

The distinctive electronic configuration, intimately tied to the quasi-1D atomic chains, yields extreme transport anisotropy. Shifting the chemical potential to the valence (−0.65 eV, Fig. 2e) or conduction (+0.65 eV, Fig. 2g) band gives rise to 2D Fermi surface contours composed of open, quasi-parallel 1D lines. Notably, this open Fermi surface topology is not confined to a single energy level but persists across broad energy windows (shaded regions, Fig. 2c) encompassing the exchange-split bands. Such topology signifies a complete absence of transverse dispersion, locking carrier group velocities strictly to the longitudinal chain direction ($y$-axis). Calculated energy-dependent conductivities quantitatively validate this behavior: for both holes (Fig. 2f) and electrons (Fig. 2h), the longitudinal conductivity ($\sigma_{yy}$) overwhelmingly dominates, while transverse conduction ($\sigma_{xx}$) vanishes. To quantify this spatial anisotropy, we evaluate the directional polarization ratio, $\delta^s = |\sigma_{yy}^s - \sigma_{xx}^s|/(\sigma_{yy}^s + \sigma_{xx}^s)$ (where, $s = \uparrow, \downarrow$). As corroborated by the shaded regions in Figs. 2f and 2h, $\delta^s$ consistently maintains a value of 1.0 across these wide operational windows. Because these frontier states are exclusively occupied by a single spin channel, the resulting charge transport achieves 100% spin polarization. Consequently, by applying an electrostatic gate voltage to shift the Fermi level, one can switch the active charge carriers (electrons/holes) and simultaneously reverse their pure spin polarization, all while preserving the 1D transport channel (see Fig. S4 in SI Section S4 for explicit carrier doping validations).

**C. Altermagnetic State and Spin-Axis-Layer Locking in the CuBr$_2$ Bilayer.** To physically realize the SALL architecture, we construct a vdW bilayer by vertically stacking two identical CuBr$_2$ monolayers with a 90° relative twist (Figs. 3a, 3b). After full structural relaxation, the system stabilizes in an A-type AFM ground state (see SI Section S5 for dynamic and magnetic stability). Crucially, the large interlayer vdW distance of 6.41 Å severely suppresses interlayer orbital overlap, thereby fulfilling the decoupled orthogonal hopping assumption of our theoretical Hamiltonian.

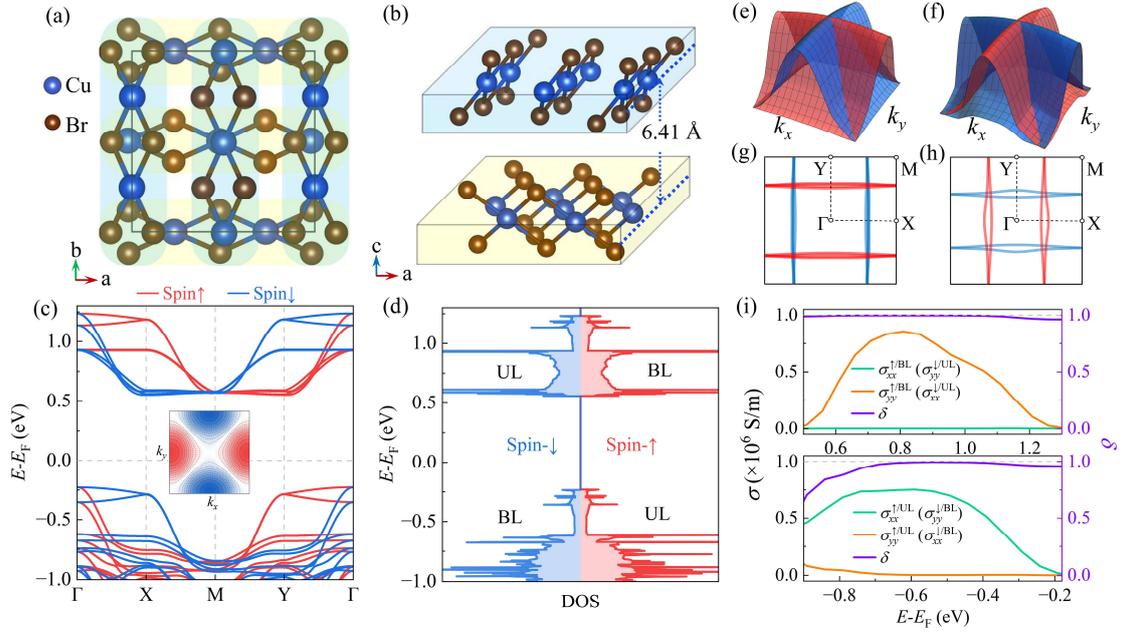

**Fig. 3.** Altermagnetic state and intrinsic SALL effect in the cross-stacked CuBr$_2$ bilayer. (a) Top and (b) perspective views of the 90°-rotated bilayer structure. (c) Spin-resolved band structure with *d*-wave spin splitting (inset). (d) Spin- and layer-projected DOS. 3D "tent-like" band dispersions for the (e) valence and (f) conduction bands. 2D orthogonal intersecting Fermi surface contours for the (g) hole- and (h) electron-doped regimes. (i) Calculated energy-dependent, spin- and layer-resolved conductivities alongside the directional polarization ratio ($\delta$) for electrons (top panel) and holes (bottom panel).

From a symmetry perspective, this cross-stacked A-type AFM configuration establishes the critical spin-group symmetry [C$_2$ ∥ C$_{4z}$] at the bilayer level. As predicted by our theoretical TB model, this symmetry operation dictates the emergence of a *d*-wave altermagnetic state, which is quantitatively corroborated by our first-principles calculations. The spin-resolved band structure (Fig. 3c) exhibits strict degeneracy along the Γ-M diagonal, alongside profound, oppositely polarized spin splittings along the orthogonal Γ-X and Γ-Y paths. The inset in Fig. 3c visualizes this momentum-dependent spin splitting ($\Delta E(\boldsymbol{k}) = E_\uparrow - E_\downarrow$) across the 2D Brillouin zone, displaying a textbook $d_{x^2-y^2}$-wave alternating pattern. Furthermore, the spin- and layer-projected DOS in Fig. 3d provides the ultimate microscopic validation for our 4×4 SALL Hamiltonian matrix. In the conduction band, spin-up states are strictly localized in the BL, whereas spin-down states reside exclusively in the UL; for the valence band, this spin-layer correspondence is exactly inverted (see SI section S6). This feature guarantees that charge doping inherently selects both the specific spin channel and the active spatial layer (axis direction).

The electronic decoupling of orthogonal 1D chains gives rise to unique "tent-like" dispersions for both the valence (Fig. 3e) and conduction (Fig. 3f) bands. Shifting the Fermi level into these tent states yields orthogonal, intersecting 1D Fermi surface contours (Figs. 3g, h). Unlike conventional 2D electron gases, these orthogonal 1D channels are fundamentally disjointed: they belong to opposite spin channels and are spatially isolated in different layers. Governed by this configuration, the corresponding charge conductivities under electron-doping (Fig. 3i, top panel) are overwhelmingly dominated by the spin-up channel in the BL along the $y$-axis ($\sigma_{yy}^{\uparrow/BL}$) and the spin-down channel in the UL along the $x$-axis ($\sigma_{xx}^{\downarrow/UL}$). The transverse conductivities for each specific spin-layer channel remain negligible, maintaining a directional polarization ratio ($\delta$) approaching 1.0 over a broad energy window. A perfectly symmetric, yet reversed, transport behavior is observed in the hole-doped regime (Fig. 3i, bottom panel). Consequently, a uniaxial in-plane electric field selectively drives a fully spin-polarized current confined entirely within a single atomic layer. By simply switching the gate voltage from n-type to p-type, the carrier polarity, spin polarization, and active transport layer are all simultaneously and reversibly switched, flawlessly realizing the intrinsic SALL device concept.

**D. All-Electrical Spin Transport and Optimal Charge-to-Spin Conversion.** The ultimate goal of the SALL architecture is to harness the highly decoupled, spin-momentum-layer locked 1D channels for functional spintronic operations. To macroscopically quantify this, we investigate the transport behavior under an arbitrary in-plane electric field. By applying a coordinate rotation transformation to the principal conductivity tensor, we calculate the effective conductivities along an arbitrary longitudinal direction ($x'$) oriented at an angle $\theta$ relative to the principal crystal axis ($x$).

Taking the electron-doped regime ($E - E_F = 0.65$ eV) as a representative case, the angular dependence of the conductivities ($\sigma_{x'x'}$ and $\sigma_{y'y'}$) for both spin channels is plotted in Fig. 4a. Governed by the orthogonal 1D nature of the constituent layers, the spin-up and spin-down conductivities exhibit characteristic $\sin^2\theta$ and $\cos^2\theta$ dependencies, respectively, resulting in a strict $\pi/2$ phase shift between the two spin channels. Consequently, the longitudinal spin polarization, $\eta = (\sigma_{x'x'}^{\uparrow} - \sigma_{x'x'}^{\downarrow})/(\sigma_{x'x'}^{\uparrow} + \sigma_{x'x'}^{\downarrow})$, can be continuously and precisely tuned from −1 to 1 simply by rotating the in-plane electric field (Fig. 4b).

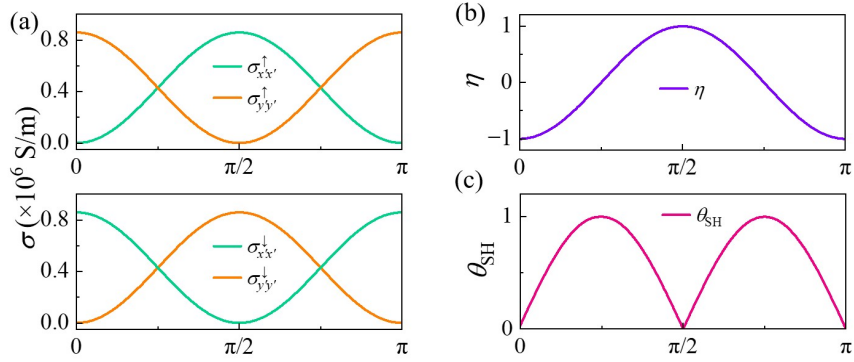

**Fig. 4.** Macroscopic anisotropic transport and optimal pure spin current generation. Angular dependence of the calculated transport properties for the electron-doped regime ($E-E_F = 0.65$ eV). (a) Longitudinal ($\sigma_{x'x'}$) and transverse ($\sigma_{y'y'}$) conductivities for the spin-up (top panel) and spin-down (bottom panel) channels as a function of the electric field angle $\theta$. (b) Tunable longitudinal spin polarization ($\eta$) driven by the in-plane electric field angle. (c) The charge-to-spin conversion efficiency (effective spin Hall angle, $\theta_{SH}$), reaching a theoretical maximum of 1.0 at $\theta = \pi/4$ and $3\pi/4$.

The most striking functional consequence emerges when the electric field is applied diagonally at $\theta = \pi/4$. At this angle, the longitudinal conductivities for both spins are exactly equal ($\sigma_{x'x'}^{\uparrow} = \sigma_{x'x'}^{\downarrow}$), rendering the longitudinal charge current completely unpolarized ($\eta = 0$). However, the transverse off-diagonal conductivities ($\sigma_{y'x'}$) for the spin-up and spin-down carriers are equal in magnitude but strictly opposite in sign. This transverse cancelation of charge, combined with the constructive accumulation of spin, generates a robust, pure transverse spin current. Importantly, unlike conventional spin Hall effects where opposite spins mix within the same physical space, the pure spin current generated in our SALL architecture is intrinsically layer-separated. Specifically, the transverse spin-up and spin-down components reside exclusively in the bottom and upper layers, respectively. This natural spatial segregation significantly facilitates the independent electrical extraction and utilization of specific spin currents via top and bottom local electrodes, bypassing the need for additional spin-sorting mechanisms. To evaluate the performance of this transverse spin generation, we compute the effective spin Hall angle ($\theta_{SH}$), defined as the ratio of the generated transverse spin conductivity to the longitudinal charge conductivity. As demonstrated in Fig. 4c, $\theta_{SH}$ reaches a theoretical maximum of 1.0 at $\theta = \pi/4$. This signifies a 100% charge-to-spin conversion efficiency—a highly coveted milestone in spintronics, achieved here intrinsically without relying on external magnetic fields, strong spin-orbit coupling, or symmetry-breaking strain. Detailed mathematical derivation of this macroscopic transport behavior, along with its equally robust manifestation in the hole-doped (p-type) regime, are provided in Section S7 and Fig. S6, Supporting Information.

In summary, we propose the spin-axis-layer locking (SALL) model to overcome reliance on external strain and valley polarization in conventional bipolar altermagnetic semiconductors. Using quasi-1D FM $CuBr_2$ monolayer as a prototype, we demonstrate that orthogonal bilayer stacking intrinsically reconstructs the [$C_2 \parallel C_{4z}$] spin-group symmetry, yielding a robust *d*-wave altermagnetic state. Severe orbital confinement decouples the orthogonal transport channels, establishing a rigid, three-way entanglement among carrier spin, transport axis, and spatial layer. Consequently, this pristine vdW platform enables simultaneous, all-electrical switching of spin polarity, carrier type, and active conduction layer via simple electrostatic gating. Furthermore, a diagonal in-plane electric field drives a transverse pure spin current with 100% charge-to-spin conversion efficiency. The intrinsic segregation of opposite spins into distinct atomic layers allows direct electrical extraction via local electrodes, bypassing conventional spin-sorting bottlenecks. Beyond CuBr$_2$, the strategy of cross-stacking 1D magnetic building blocks provides a generalizable framework for exploring coupled degrees of freedom in completely intrinsic, stimuli-free platforms, advancing the realization of highly integrated spintronic devices.

**Acknowledgments.** This work is supported by the National Natural Science Foundation of China (12464040), the Natural Science Foundation of Inner Mongolia Autonomous Region (2021JQ-001), and the 2020 Institutional Support Program for Youth Science and Technology Talents in Inner Mongolia Autonomous Region (NJYT-20-B02).